\documentstyle{lamuphys}
\input epsf
\epsfverbosetrue
\newlength{\ppindent}
\def \be {\begin{equation}}
\def \ee {\end{equation}}
\def \sgn {{\rm sgn}}

\def \la{\left(}
\def \ra{\right)}
\def \ld{\left\langle}
\def \rd{\right\rangle}
\def \ldd{\left\langle\!\left\langle}
\def \rdd{\right\rangle\!\right\rangle}
\def \cP{{\cal P}}
\def \cF{{\cal F}}
\def \cN{{\cal N}}
\def \cZ{{\cal Z}}

\def \vW{\vec{W}}

\def \vnu{\vec{\nu}}
\def \vxi{\vec{\xi}}
\def \vzeta{\vec{\zeta}}
\begin{document}
\title{Neural Networks}
\author{Heinz Horner and Reimer K\"uhn}
\institute{Institut f\"ur Theoretische Physik, Universit\"at Heidelberg\\
Philosophenweg 19, 69120 Heidelberg, Germany}
\date{1. Nov. 1996} 
\maketitle
%
%
\begin{abstract}
We review the theory of neural networks, as it has emerged in the last ten
years or so within the physics community, emphasizing  questions of biological 
relevance over those of importance in mathematical statistics and machine learning theory.
\end{abstract}
%
\section{Introduction\label{sec1}}
Understanding at least some of the functioning of our own brain is
certainly an extraordinary scientific and intellectual challenge and
it requires the combined effort of many different disciplines. Each
individual group can grasp only a limited set of aspects, but
its particular methods, questions and results can influence,
stimulate and hopefully enrich the thoughts of others. This is the
frame in which the following contribution, written by theoretical
physicists, should be seen.

Statistical physics usually deals with large collections of similar
or identical building blocks, making up a gas, a liquid or a solid.
For the collective behavior of such an assembly most of the properties
of the individual elements are only of marginal relevance. This
allows to construct crude and simplified models which nevertheless
reproduce certain aspects with extremely high accuracy. An essential
part of this modeling is to find out which of the properties of the
elements are relevant and what kind of questions can or cannot be
treated by such models. The usual goal is to construct models as simple
as possible and to leave out as many details as possible, even if they
are perfectly well known. The natural hope is that the essential
properties can be understood better on a simple model. This, on the
other hand, seems to contrast the ideals of modeling in other
disciplines and this can severely obstruct the interdisciplinary
exchange of thoughts.

Our brain (\cite{Brai}) is certainly not an unstructured collection of
identical neurons. It consists of various areas performing special tasks
and communicating along specific pathways. Even on a smaller scale it is
organized into layers and columns. Nevertheless the overwhelming
majority of neurons in our brain belongs to one of perhaps three
types. Furthermore, on an even smaller scale, neurons seem to interact
in a rather disordered fashion, and the pathways between different
areas are to some degree diffuse. Keeping in mind that models of
neural networks with no a priori structure are certainly limited, it is
of interest to see how structures can evolve by learning processes
and what kind of tasks they can perform.

Over the last ten or more years, abstract and simplified models of
brain functions have been a target of research in statistical
physics (\cite{Amit}; \cite{HKP}). A model of an associative memory was
proposed by Hopfield in 1982, following earlier work by Caianello and
Little. This model is not only based on extremely simplified neurons
(McCulloch-Pitts, 1943), it also serves a heavily schematized task, the
storage and retrieval of uncorrelated random patterns. This twofold
idealization made it, however, tractable and accessible for quantitative
results. In the meantime there have been many extensions of this model,
some of which will be discussed later on. One of the essential points of
this model is the fact that information is stored in a distributed fashion
in the synaptic connections among the neurons. Each synapse carries
information about each pattern stored, such that destruction of part
of the synapses does not destroy the whole memory. The storage of a
pattern requires a learning process which results in a modification
of the strength of all synapses. The original model was based on a
simple learning rule, essentially the one proposed by Hebb, which is in
a sense a neuronal manifestation of Pawlow's ideas of conditional
reflexes. Regarding learning, again more sophisticated rules have been
investigated and are discussed later.

Even restricting ourselves to this kind of models, we can sketch only a
small part of what has been worked out in the past, and only small parts
also of the progress in getting those models closer to biology. It is 
interesting to note that artificial neural nets, in the form of algorithms 
or hardware, have found many technical applications. This aspect will,
however, be left aside almost completely.

Before entering the discussion of learning or memory, we want to give
a brief overview over the biological background of neurons, their
basic functioning and their arrangement in the brain (\cite{Brai}; 
\cite{Abe}).


\section{Biological vs. Formal Neural Nets\label{sec2}}
\subsection{Biological Background}

A typical neuron, e.g. a pyramidal cell (see Fig.~1 next page),
consists of the cell body or soma; extending from it there is a branched
structure  of about 2 mm diameter, called dendrite, and the nerve fiber
or axon, which again branches and can have extensions reaching distant
parts of the brain. The branches of the axon end at so called synapses
which make contact to the dendrites of other neurons. There are of
course also axons coming in from sensory organs or axons reaching out to
the motor system. Compared to the number of connections within the
brain, their number is rather small. This amazing fact indicates perhaps
that the brain is primarily busy analyzing the sparse input or shuffling
around internal information.

The main purpose of a neuron is to receive signals from other
neurons, to process the signals and finally to send signals again to other cells. What happens in more detail is the following. Assume a cell is 
excited, which means that the electrical potential across its membrane 
exceeds some threshold. This creates a short electric pulse, of about 1 
msec duration, which travels along the axon and ultimately reaches the 
synapses at the ends of its branches. Having sent a spike, the cell returns 
to its resting state. A spike arriving at a synapse releases a certain amount 
of so called neuro--transmitter molecules which diffuse across the small gap 
between synapse and dendrite of some other cell. The neuro--transmitters
themselves then open certain channel proteins in the membrane
%
%
%
of the postsynaptic cell and this finally influences the electrical 
potential across the membrane of this cell. The neuro--transmitters 
released from pyramidal cells have the effect of driving
the potential of the postsynaptic cell towards the threshold, their
synapses are called excitatory. There are, however, also inhibitory
cells with neuro--transmitters having the opposite effect. The individual
changes of the potential caused by the spikes of the presynaptic cells
are collected over a period of about 10 msec and if the threshold is
reached the postsynaptic cell itself fires a spike. Typically 100 
incoming spikes within this period are necessary to reach this state.

\begin{figure}[h]
\centering
\framebox[7cm]{
\leavevmode
  \epsfysize=4cm
  \epsffile{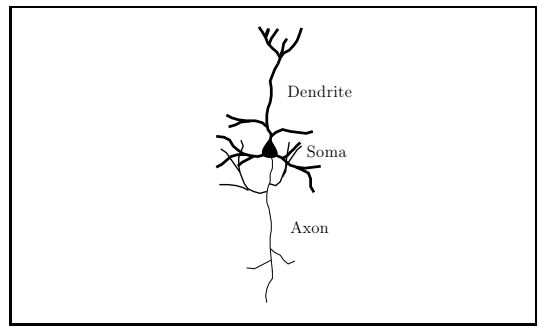}
}
\caption{Schematic view of a neuron}
\end{figure}

The human brain contains $10^{10}$ to $10^{11}$ neurons and more than
$10^{14}$ synaptic connections among them. The neurons are arranged in
a thin layer of about 2 mm on the surface, the cortex, and each mm$^2$
contains typically $10^5$ such cells. This means that the dendritic
trees of these cells penetrate each other and form a dense web. Part
of the axons of these cells again project onto the dendrites in the
immediate neighborhood and only a fraction reaches more distant
regions of the brain. This means that on a scale of a few mm$^3$
more than $10^5$ neurons are tightly connected. This does not imply
that more distant regions are weakly coupled. The huge amount of
white matter containing axons connecting more distant parts only
indicates the possibility of strong interactions of such regions as
well.

It is tempting to compare this with structures which we find within
the integrated circuits of an electronic computer. The typical size
of a synapse is 0.1 $\mu$m, whereas the smallest structures found in
integrated circuits are about five times as big. The packing density
of synapses attached to a dendrite is about $10^9$ per mm$^2$,
whereas only $1/1000$ of this packing density is reached in electronic
devices. A comparison of the computational power is also impressive. A
modern computer can perform up to $10^9$ elementary operations per
sec. The computational power of a single neuron is rather low, but
they all work in parallel. Assuming that a neuron fires with a rate
of 10 spikes per sec, which is typical, and assuming that each spike
transmitted through a synapse corresponds to an elementary
computation, we find a computing power of about $10^{15}$ operations
per second. These numbers have to be kept in mind if we try to imitate
brain functions with artificial devices.

\subsection{Formal Neurons --- Spikes vs. Rates}

The actual processes going on when a spike is formed or when it
arrives at a synapse and its signal is transmitted to the next
neuron, involve an interplay of various channels, ionic currents
and transmitter molecules. This should not be of concern as long
as we are interested only in the data processing aspects. A
serious question is, however, what carries the information? Is it a
single spike and its precise timing or is the information coded in
the firing rates? For sensory neurons the proposition of rate coding
seems well established. These neurons typically have rather high
firing rates in their excited state. For the brain this is much less
clear since the typical spike rates are low and the intervals
between two successive spikes emitted by a neuron are longer or at
best of the order of the time over which incoming spikes are
accumulated. Nevertheless a spike rate coding is usually assumed for
the brain as well. This means that a rate has to be considered as
an average over the spikes of many presynaptic neurons rather than a
temporal average over the spikes emitted by a single cell. This is
plausible, having in mind that typically 100 or more arriving spikes
are necessary to release an outgoing spike. This suggests for the
firing rate $\nu_i$ of a neuron $i$
\begin{equation}\label{2a}
\nu_i(t)=\Phi\left(\sum_j W_{ij}\nu_j(t)-\vartheta\right)\ ,
\end{equation}
where $\vartheta$ is some threshold and $\Phi(x)$ is some increasing
function of $x$. The simplest assumption is $\Phi(x)=0$ for $x<0$
and $\Phi(x)=1$ for $x>0$. This model was first proposed by 
McCulloch and Pitts (1943). The quantity $W_{ij}$ describes the coupling
efficacy of the synapse connecting the presynaptic cell $j$ with the
postsynaptic cell $i$. For excitatory synapses $W_{ij}>0$, and for
inhibitory synapses $W_{ij}<0$.

This model certainly leaves out many effects. For instance the assumption
of linear superposition of incoming signals neglects any dependence on
the position of the synapse on the dendritic tree.

Investigating networks of spiking neurons, one has again designed 
simplified models. One of them is the integrate and fire neuron which
mimics the mechanism of spike generation at least in a crude way. It
sums up incoming signals by changing the membrane potential. As soon as
a certain threshold is reached, the neuron fires and the membrane
potential is reset to its resting value. 

If rate coding is appropriate, results obtained for the first kind of
networks should be reproduced by networks of spiking neurons as well.
On the other hand there are many questions which can only be taken up
within the framework of spiking neurons, for instance which role the
precise timing of spikes plays (\cite{Abe}) or whether the activity in a
network with excitatory couplings among its neurons can be stabilized by
adding inhibitory neurons (\cite{AB}).

\subsection{Hebbian Learning --- Sparse Coding}

The most remarkable feature of neural networks is their ability 
to learn. This is attributed to a certain plasticity of the
synaptic coupling strengths. The question is of course, how is
this plasticity used in a meaningful way?

The basic idea proposed by Hebb (1949) actually goes back to the notion
of conditional reflexes put forward by Pawlow. Assume a stimulus
$A$ results in a reaction $R$. If simultaneously with $A$ a second
stimulus $B$ is applied, then after some training stimulus $B$ alone will
be sufficient to trigger reaction $R$, although this was not the case
before training. Let $A$ be represented by the activity of a
neuron $\ell$ and the reaction $R$ by neuron $i$ becoming active.
This would be the case if the coupling $W_{i\ell}$ is sufficiently
strong. Before training, the stimulus $B$, represented by the
activity of neuron $j$, is assumed {\em not\/} to trigger the reaction $R$.
That is, the coupling $W_{ij}$ is assumed to be weak. During training
with $A$ {\em and\/} $B$ present, the cells $j$ and $i$ are
simultaneously active, the latter being activated by cell $\ell$.
Assume now that the synaptic strength $W_{ij}$ between 
neurons $j$ and $i$ is increased, if both cells are
simultaneously active. Then, after some training, this coupling
$W_{ij}$ will be strong enough to sustain the reaction $R$
without $A$ being applied, provided $B$ is present. This is
represented by the Hebb learning rule
\begin{equation}\label{2b}
\Delta W_{ij} \propto \nu_i\,\nu_j
\end{equation}
Most remarkably this learning rule does not require a direct
connection between the cells $\ell$ and $j$ representing the stimuli
$A$ and $B$. That is, the equivalence of stimuli $A$ and $B$
has been learnt without any a priori relation between $A$ and
$B$. What has been used is only the simultaneous occurrence of $A$
and $B$. Despite its simplicity this learning rule is extremely
powerful.

It is not completely clear how such a change in the synaptic
efficacies is realized in detail, whether it is caused by changes
in the synapse itself or by changes in the density of receptor
proteins on the membrane of the dendrite of the postsynaptic
neuron. Nevertheless it is plausible at least in the sense that
this learning process depends only on the simultaneous state of the
pre- and postsynaptic cell. It is generally assumed that learning
takes place on a time scale much slower than the intrinsic time
scale of a few msec characteristic of neural dynamics.

We can now go one step further and consider the learning of more
than one pattern. A pattern is a certain configuration of active
and inactive neurons. A pattern, say $\mu$, is represented by a
set of variables $\xi_i^\mu$ for each pattern and neuron. This
means that in pattern $\mu$ neuron $i$ fires with a rate
$\xi_i^\mu$. In the most simple case $\xi_i^\mu=1$ if $i$ is
excited and $\xi_i^\mu=0$ otherwise. Having learnt a set of
patterns the couplings, according to the above learning rule, have
the values
\begin{equation}\label{2c}
W_{ij}=W_o\sum_\mu \xi_i^\mu\,\xi_j^\mu
\end{equation}
where $W_o$ has the meaning of a learning strength. Actually this
learning strength might also depend on the kind of pattern
presented, for instance on whether the pattern is new, unexpected,
relevant in some sense or under which global situation, attention,
laziness or stress, it is presented. This can lead to improved
learning or suppression of uninteresting information.

The above learning rule is constructed such that, at least for
$W_o>0$, only excitatory couplings are generated. This is in
accordance with the finding that the pyramidal neurons have
excitatory synapses only and that the plasticity of the synapses
is most pronounced in this cell type. This causes, on the other
hand, a problem. A network with excitatory synapses only would
shortly go into a state where all neurons are firing at a high
rate. The cortex contains, however, inhibitory cells as well. The
likely purpose of these cells is to control the mean activity of
the network and to prevent it from reaching the unwanted state of
uniform high activity. A malfunctioning of this regulation is
probably the cause of epileptic seizures.

Actually the mean activity in our brain seems to be rather low. This
means that at a given time only a small percentage of the neurons
is firing at an elevated rate. Typical patterns are
sparse, having many more $0$'s than $1$'s. This is a bit surprising
since the maximal information per pattern is contained in
binary patterns with approximately equal number of  $0$'s and
$1$'s and such symmetric coding is also used in our computers.
Nevertheless there are several good reasons for sparse coding,
some of which will be discussed later.

In the original Hopfield model the degree of abstraction is pushed
a step further. Here symmetric patterns with equal number of $0$'s
and $1$'s are considered. This requires a modified learning rule.
First of all the inactive state is now represented by $-1$ rather
than $0$. With this modification the above learning rule can again
be used, but now the coupling strength can also be weakened and the
couplings can acquire negative values. Furthermore it is assumed
that each neuron is connected to every other neuron and that
the couplings between two neurons have the same value in both
directions. This is certainly rather unrealistic in view of the
biological background. Modified models with one or the other
simplifying assumption removed have been investigated as well. They 
show, however, quite similar behavior. This demonstrates the
robustness of the models with respect to modifications of details,
which might again serve as a justification of this simple kind of
modeling.

\subsection{Transmission Delays}

The propagation of a spike along the axon, the transmission
of this signal across the synapse and the propagation along the
dendrite take some time. This causes some total delay $\tau_{ij}$,
typically a few msec, in the transmission of a signal from neuron
$j$ to neuron
$i$. Incorporating this into eq.(\ref{2a}) yields a modified form
\begin{equation}\label{2d}
\nu_i(t)=\Phi\left(\sum_j W_{ij}\nu_j(t-\tau_{ij})-\vartheta\right).
\end{equation}

As long as we are interested in slow processes, this delay is of
no relevance. On the other hand it gives the opportunity to
generate or learn sequences of patterns evolving in time. This
might be of relevance in processes like speech generation or
recognition, or in generating periodic or aperiodic motions. Other
proposals use this mechanism for temporal linking of different
features of the same object or for the segmentation of stimuli
generated by unrelated objects. Another mechanism which might
play a role in this context is the phenomenon of fatigue. This
means that the firing rate of an excited neuron, even at constant
input, goes down after a while. The associated time scales can
vary from few msec up to minutes or hours. In any case there are
several mechanisms which can be used for the generation or
recognition of temporal structures and we are coming back to this
point later.

The picture developed so far certainly leaves out many interesting
and important aspects. Nevertheless even this oversimplified frame
allows to understand some basic mechanisms. On the other hand it is
far from a description of the brain as a whole. What is certainly 
missing, is the structure on a larger scale. In order to
proceed in this direction one would have to construct modules
performing special tasks, like data preprocessing or memory, and
one would have to arrange for a meaningful interplay of those
modules. This is currently far beyond our possibilities, as we are 
lacking analytic tools or computational power and, perhaps more 
importantly even, good questions and well formalizable tasks to be 
put to such a modular architecture.


\section{Learning and Generalization\label{sec3}}

Given that we interpret the firing patterns of a neural network as representing
information, neural dynamics must be regarded as a form of {\em information 
processing}. Moreover, disregarding the full complexity of the internal dynamics of single neurons, as we have good reasons to do (see Sec. 2.2), we find 
the course of neural dynamics, hence information processing in a neural networks, being determined by its synaptic organization. 

Consequently, {\em shaping\/} the information processing capabilities of a
neural network requires changing its synapses. In a neural setting, this process
is called  ``learning", or ``training", as opposed to ``programming" in the
context of symbolic computation. Indeed, as we have already indicated above, the
process of learning is rather different from that of programming a computer.
It is incremental, sometimes repetitive, and it proceeds by way of presenting 
``examples". The examples may represent associations to be implemented in the
net. They may also be instances of some rule, and one of the reasons for 
excitement about neural networks is that they are able to {\em extract\/} rules 
from examples. That is, by a process of training on examples they can be made 
to {\em behave\/} according to a set of rules which --- while manifest in the 
examples --- are usually never made explicit, and are quite often not known in 
algorithmic detail. Such is, incidentally, also the case with most skills humans
possess (subconcsiously). In what follows, we discuss the issues of learning
and generalization in somewhat greater detail.

We start by analysing learning (and generalization) for a single threshold 
neuron, the perceptron. First, because it gives us the opportunity to discuss 
some of the concepts useful for a quantitative analysis of learning already in 
the simplest possible setting; second, because the simple perceptron can be 
regarded as the elementary building block of networks exhibiting more 
complicated architectures, and capable of solving more complicated tasks.

\begin{figure}[h]
\centering
\framebox[7cm]{
\setlength{\unitlength}{0.4cm}
\begin{picture}(12,8)
	\put(-2,0){
		\begin{picture}(11,6)(-1,-1.5)
		\thicklines
		\multiput(0,0)(3,0){3}	{\circle*{0.5}}
		\multiput(1,2)(2,0){3}	{\circle*{0.5}}
		\put(3,4)		{\circle*{0.5}}
		\put(0,0) {\vector(1,2) {0.9}}
		\put(0,0) {\vector(3,2) {2.8}}
		\put(3,0) {\vector(-1,1){1.82}}
		\put(3,0) {\vector(0,1) {1.75}}
		\put(3,0) {\vector(1,1) {1.82}}
		\put(6,0) {\vector(-3,2){2.8}}
		\put(6,0) {\vector(-1,2){0.9}}
		\put(1,2) {\vector(1,1) {1.82}}
		\put(3,2) {\vector(0,1) {1.75}}
		\put(5,2) {\vector(-1,1){1.82}}
		\put(3,4) {\vector(0,1) {1.5}}
		\put(3,-1){\makebox(0,0){(a)}}
		\put(7.5,0){
			\put(3,0) {\circle*{0.5}}
			\put(1,1) {\circle*{0.5}}
			\put(5,3) {\circle*{0.5}}
			\put(1,4) {\circle*{0.5}}
			\put(3,5) {\circle*{0.5}}
			\put(3,0) {\vector(2,3)  {1.8}}
			\put(3,0) {\vector(0,1)  {4.8}}
			\put(3,0) {\vector(-2,1) {1.8}}
			\put(3,0) {\vector(-1,2) {1.9}}
			\put(5,3) {\vector(-2,-3){1.8}}
			\put(5,3) {\vector(-2,-1){3.8}}
			\put(5,3) {\vector(-4,1) {3.8}}
			\put(5,3) {\vector(-1,1) {1.8}}
			\put(3,5) {\vector(0,-1) {4.8}}
			\put(3,5) {\vector(1,-1) {1.8}}
			\put(3,5) {\vector(-2,-1){1.8}}
			\put(1,1) {\vector(0,1)  {2.8}}
			\put(1,1) {\vector(2,1)  {3.8}}
			\put(3,-1){\makebox(0,0) {(b)}}
			}
		\end{picture}}
\end{picture}}
\caption{{\bf (a)} Feed--forward Network. {\bf (b)} Network with feed--back 
loops}
\end{figure}
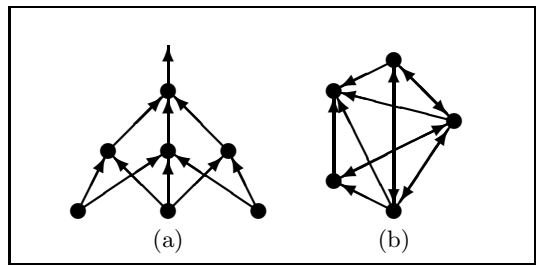

Regarding architectures, it is useful to distinguish between so called
feed--forward nets, and networks with feedback--loops (Fig.~2). In feed--forward
nets, the information flow is directed; at their output side, they produce a certain {\em map} or function of the firing patterns fed into their input layer.
Given the architecture of such a layered net, the function it implements is determined by the values of the synaptic weights between its neurons. Networks with feedback--loops, on the other hand exhibit and utilize non--trivial 
{\em dynamical\/} properties. For them, the notion of (dynamical) attractor is 
of particular relevance, and learning aims at constructing desired attractors, 
be they fixed points, limit cycles or chaotic. We discuss attractor networks 
separately later on in Sec. 4. Finally, feed--forward architectures 
may be combined with elements providing feedback--loops in special ways to 
create so--called feature maps, which we also briefly describe. 

The physics--approach to analyzing learning and generalization has consisted 
in supplementing general considerations with quantitative analyses of heavily 
schematized situations. Main tools have been statistical analyses, which can
however be quite forceful (and luckily often simple) when the size of a given
information processing task becomes large in a sense to be specified below. 

It goes without saying that this approach would not be complete without
demonstrating --- either theoretically, by way of simulations, or, by studying 
special examples --- that the main functional features and trends seen in 
abstract statistical settings would survive the removal of a broad range
of idealizations and simplifications, and that they, indeed, prove to be resilient against changing fine details at the microscopic level.

\subsection{Simple Perceptrons}

A perceptron mimics the functioning of a single (formal) neuron. Given an 
input $\vnu = (\nu_1,\nu_2,\dots ,\nu_N)$ at its $N$ afferent synapses, it 
evaluates its local field or post-synaptic potential as weighted sum of the 
input components $\nu_j$, 
\be
h_0(\vnu) = \sum_{j=1}^N W_{0j} \nu_j\ ,
\label{P1}
\ee
compares this with a threshold $\vartheta$, and produces an output $\nu_0$ according to its transfer function or input--output relation
\be
\nu_0 = \Phi(h_0(\vnu) - \vartheta)\ ,
\label{P2}
\ee
For simple perceptrons, one usually assumes a step--like transfer function.
Common choices are $\Phi(x) = \sgn(x)$ or $\Phi(x)=\Theta(x)$ depending on whether one chooses a $\pm 1$ representation or a 1--0 representation for the 
active and inactive states\footnote{$\Theta(x)$ is Heaviside's step function: 
$\Theta(x)=1$ for $x > 0$ and $\Theta(x) = 0$ otherwise.}.

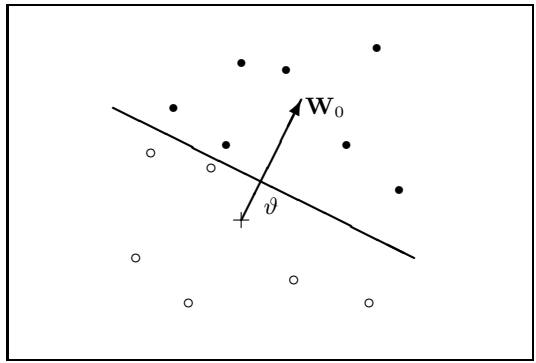
\begin{figure}[h]
\centering
\framebox[7cm]{
\setlength{\unitlength}{1cm}
\begin{picture}(5,4.5)
	\put(0,0){
		\begin{picture}(4,4)(-2.25,-2.25)
		\thicklines
		\put(-2,1)	{\line(2,-1){4}}
		\put(-0.3,-0.5)	{\makebox(0,0){$+$}}
		\put(-0.3,-0.5)	{\vector(1,2){0.8}}
		\thinlines
		\put(1.1,0.5)	{\circle*{0.1}}
		\put(1.8,-0.1)	{\circle*{0.1}}
		\put(1.5,1.8)	{\circle*{0.1}}
		\put(0.3,1.5)	{\circle*{0.1}}
		\put(-1.2,1)	{\circle*{0.1}}
		\put(-0.3,1.6)	{\circle*{0.1}}
		\put(-0.5,0.5)	{\circle*{0.1}}
		\put(-1,-1.6)	{\circle{0.1}}
		\put(-1.7,-1)	{\circle{0.1}}
		\put(-1.5,0.4)	{\circle{0.1}}
		\put(-0.7,0.2)	{\circle{0.1}}
		\put(0.4,-1.3)	{\circle{0.1}}
		\put(1.4,-1.6)	{\circle{0.1}}
		\put(0,-0.3)	{\makebox(0,0)[l]{$\vartheta$}}
		\put(0.55,1)	{\makebox(0,0)[l]{$\vec{W}_0$}}
		\end{picture}}
\end{picture}
}
\caption{Linear separation by a perceptron}
\end{figure}

The kind of functionality provided by a perceptron has a simple geometrical
interpretation. Equation (\ref{P2}) shows that a perceptron implements
a {\em two-class\/} classification, assigning an `active' or an `inactive' 
output--bit to each input pattern $\vnu$, according to whether it produces a 
super- or sub--threshold local field. The dividing {\em decision surface\/} 
is given by the inputs for which $\vW_0\cdot\vnu \equiv \sum_j W_{0j} \nu_j =
\vartheta$. It is a {\em linear\/} hyperplane orthogonal to the direction of
the vector $\vW_0$ of synaptic weights $(W_{0j})$ in the $N$-dimensional space 
of inputs (Fig.~3). Pattern sets which are classifiable that way are called 
linearly separable. The linearly separable family of problems is certainly 
non--trivial, but obviously also of limited complexity. Taking Boolean functions of two inputs as an example, and choosing the representation $1 \equiv {\em 
true}$ and $0 \equiv {\em false}$, one finds that ${\rm AND}(\nu_1,\nu_2)$, 
and ${\rm OR}(\nu_1,\nu_2)$ as well as ${\rm IMPL}(\nu_1,\nu_2)$ are linearly 
separable, whereas ${\rm XOR}(\nu_1,\nu_2)$ is not. 

It is interesting to see, how Hebbian learning, the most prominent candidate
for a biologically plausible learning algorithm, would perform on learning
a linearly separable set of associations. A problem that has been thoroughly
studied is that of learning {\em random\/} associations. That is, one is given 
a set of input patterns $\vxi^\mu$, $\mu=1,\dots ,P$, and their associated set of desired output labels $\zeta_0^\mu$. Each bit in each pattern is 
independently chosen to be either active or inactive with equal probability 
and the same is assumed for the output bits.

It has been known for some time (\cite{cov}) that such a set of random 
associations is {\em typically\/} linearly separable, as long as the number 
$P$ of patterns does not exceed twice the dimension $N$ of the input space, 
$P\le 2 N$. It turns out that the suitable representation of the active and inactive states for this problem --- i.e., appropriate for the given pattern
statistics --- is a $\pm 1$ representation. Moreover, due to the symmetry 
between active and inactive states in the problem, a zero threshold should
be chosen.

Learning \`a la Hebb by correlating pre- and postsynaptic activities, one 
has $(\Delta W_{0j})^\mu \propto \zeta_0^\mu \xi_j^\mu$ as the synaptic 
change in response to a presentation of pattern $\mu$. As we have mentioned 
already, this involves a modification of Hebb's original proposal. Summing 
contributions from all patterns of the problem set, one obtains (compare
Eq.~(3))
\be
W_{0j} = \frac{1}{N} \sum_{\mu=1}^P \zeta_0^\mu \xi_j^\mu\ ,
\label{P3}
\ee
where the prefactor is chosen just to fix scales in a manner that allows 
taking a sensible large system limit. Here we distinguish input from output
bits by using different symbols for them. In recursive networks, outputs of
single neurons are used as inputs by other neurons of the same net, and the
distinction will be dropped in such a context.

It is not difficult to demonstrate that Hebbian learning finds an {\em
approximation\/} to the separating hyperplane, which is rather good for small
problem size $P$, but which becomes progressively worse as the number of
patterns to be classified increases. To wit, taking an arbitrary example
$\vxi^\nu$ out of the set of learnt patterns, one finds that the Hebbian 
synapses (\ref{P3}) produce a local field of the form $h^\nu=h_0(\vxi^\nu) = 
\zeta_0^\nu + \delta^\nu$. Here $\zeta_0^\nu$ is the correct output-bit 
corresponding to the input pattern $\vxi^\nu$ (the {\em signal}), which is 
produced by the $\nu$--th contribution  $(\Delta W_{0j})^\nu$ to the $W_{0j}$. 
The other contributions to $h^\nu$ do not add up constructively. Together they 
produce the {\em noise} term $\delta^\nu$. In the large system limit, one can 
appeal to the central limit theorem to show that the probability density of 
the noise is Gaussian with zero mean and variance $\alpha = P/N$.\footnote{The
precise value is actually $(P-1)/N$.} A misclassification occurs, if the noise 
succeeds in reversing the sign determined by the signal $\zeta_0^\nu$. Its 
probability depends therefore only on $\alpha$, the ratio of problem size $P$ 
and system size $N$. It is exponentially small --- $P_{\rm err}(\alpha) \sim 
\exp(-1/2\alpha)$ --- for small $\alpha$, but increases to sizeable values
already way below $\alpha_c = 2$, which is the largest value for which the
problem is linearly separable, i.e. the largest value for which we know that a solution with $P_{\rm err}=0$ typically exists. If, however, a finite fraction of errors is tolerable, and such can be the case, when one is interested in the overall output of a large array of perceptrons, then moderate levels of loading can, of course, be accepted. We shall see in Sec.~4 below that this is a standard situation in recursive networks.

The argument just presented can be extended to show that even distorted 
versions of the learnt patterns are classified correctly with a reasonably  small error probability, provided the distortions are not too severe and, 
again, the loading level $\alpha$ is not too high. 

The modified Hebbian learning prescription may be generalized to handle low 
activity data, i.e. patterns with unequal proportions of active and inactive
bits. The appropriate learning rule is most succinctly formulated in terms of
a 1-0 representation for the active and inactive states and reads 
\be
(\Delta W_{0j})^\mu \propto \tilde\zeta_0^\mu \tilde\xi_j^\mu
\label{P4}
\ee
where $\tilde\zeta_0^\mu = \zeta_0^\mu - a_{\rm out}$ and $\tilde\xi_j^\mu 
=\xi_j^\mu - a_{\rm in}$, with $a_{\rm in/out}$ denoting the probability of 
having active bits at the input and output sides, respectively. Non-zero 
thresholds are generally needed to achieve the desired linear separation. 
Interestingly this rule ``approaches" Hebb's original prescription in the low 
activity limit $a_{\rm in/out} \to 0$; the strongest synaptic changes occur, 
if both, presynaptic and postsynaptic neuron are active, and learning 
generates predominantly excitatory synapses. Interestingly also, this rule benefits from low activity at the output side: The variance of the noise contribution to local fields is reduced by a factor $a_{\rm out}(1-a_{\rm 
out})/(1-a_{\rm in})$ relative to the case $a_{\rm in}=a_{\rm out}=\frac{1}{2}$, leading to reduced error rates and correspondingly enlarged storage capacities.
We shall return to this issue in  Sec.~4 below. 

Two tiny modifications of the Hebbian learning rule (\ref{P3}),(\ref{P4}) serve 
to boost its power considerably. First, synapses are changed in response to a 
pattern presentation only, if the pattern is currently misclassified. If $\zeta_0^\mu$ is the desired output bit corresponding to an input pattern
$\vxi^\mu$ which is currently misclassified, then
\be
(\Delta W_{0j})^\mu  \propto \varepsilon^\mu \zeta_0^\mu \xi_j^\mu\ ,
\label{P5}
\ee
where $\varepsilon^\mu$ is an error mask that signifies whether the pattern 
in question is currently misclassified ($\varepsilon^\mu = 1$) or not ($\varepsilon^\mu = 0$). Here, a $\pm 1$ representation for the output bits is 
assumed; the input patterns can be chosen arbitrarily in $\bbbr^N$. Second, 
pattern presentation and (conditional) updating of synapses according to 
(\ref{P5}) is continued as long as errors in the pattern set occur. The 
resulting learning  algorithm is called {\em percepton learning}.

An alternative way of phrasing (\ref{P5}) uses the output error $\delta_0^\mu = \zeta_0^\mu - \nu_0^\mu$, i.e., the difference between the desired $\zeta_0^\mu$ and the current actual output bit $\nu_0^\mu$ for pattern $\mu$. This gives 
$(\Delta W_{0j})^\mu \propto \delta_0^\mu\ \xi_j^\mu$. It may be read as a 
combined process of learning the desired association and ``unlearning" the 
current erroneous one. 

With Hebbian learning, perceptron learning shares the feature that synaptic changes are determined by data locally available to the synapse --- the values 
of input and (desired) output bits. Both, the {\em locality}, and the {\em 
simplicity\/} of the essentially Hebbian correlation--type synaptic updating 
rule must be regarded prerequisites for qualifying perceptron learning --- indeed {\em any\/} learning rule --- to be considered as a ``reasonable 
abstraction" of a biological learning mechanism. 

Unlike Hebbian learning proper, perceptron learning requires a supervisor or 
teacher to compare current and desired performance. Here --- as with any other
supervised learning algorithm --- is, perhaps, a problem, because neither do 
our synapses know about our higher goals, nor do we have immediate or deliberate control over our synaptic weights. It is conceivable though that the necessary 
supervision and feedback be provided by other neural modules, provided that the 
output of the perceptron in question is ``directly visible" to them and a more 
or less direct neural pathway for feedback is available. We will have occasion 
to return to this issue later on.

The resulting advantage of supervised perceptron learning over simple Hebbian 
learning is, however, dramatic. Perceptron learning is {\em guaranteed\/} to 
find a solution to a learning task after {\em finitely\/} many updatings, 
provided only that a solution exists, and no assumptions concerning pattern statistics need be made. Morevoer, learning of thresholds can, if necessary, be 
easily incorporated in the algorithm. This is the content of the so--called 
perceptron convergence theorem (\cite{ros}). For a precise formulation and for 
proofs, see (\cite{ros}; \cite{mipa}; \cite{HKP}).

So far, we have discussed the problem of storing, or embedding a set of 
(random) associations in a perceptron. It is expedient to distinguish this problem from that of {\em learning a rule\/}, given only a set of examples representative of the rule.

For the problem of learning a rule, a new issue may be defined and studied,
viz. that of {\em generalization}. Generalization, as opposed to memorization, 
is the ability of a learner to perform correctly with respect to the rule in situations (s)he has not encountered before during training.

For the perceptron, this issue may be formalized as follows. One assumes that
a rule is given in terms of some unknown but fixed separating hyperplane 
according to which all inputs are to be classified. A set of $P$ examples, 
\be
\zeta_0^\mu = \sgn(\vW^{\rm t} \cdot \vxi^\mu)\ ,\quad \mu=1,\dots,P\ ,
\label{G1}
\ee
is produced by a ``teacher perceptron", characterized by its coupling vector $\vW^{\rm t} = (W^{\rm t}_1,W^{\rm t}_2,\dots,W^{\rm t}_N)$ which represents
the separating hyperplane (the rule) to be learnt. That is, as before, the 
input patterns $\vxi^\mu$ are randomly generated; however, the corresponding 
outputs are now no longer independently chosen at random, but fixed functions 
of the inputs. A ``student perceptron" attempts to learn this set of examples 
--- called the {\em training set} --- according to some learning algorithm.

The generalization error $\varepsilon_{\rm g}$ is the probability that student 
and teacher disagree about the output corresponding to a randomly chosen input 
that was not part of the training set. For perceptrons there is a very simple
geometrical visualization for the probability of disagreement between teacher
$\vW^{\rm t}$ and student $\vW$. It is just $\varepsilon_{\rm g} = \theta/\pi$, 
where $\theta$ is the angle between the teacher's and the student's coupling vector (see Fig.~4).

\begin{figure}[t]
\centering
\framebox[7cm]{
\setlength{\unitlength}{1cm}
\begin{picture}(5,4.5)
	\put(0,0){
		\begin{picture}(4,4)(-2.25,-2)
		\thicklines
		\put(0,0)	{\line(4,-1){3}}
		\put(0,0)	{\line(-4,1){3}}
		\put(0,0)	{\line(2,-1){2.75}}
		\put(0,0)	{\line(-2,1){2.75}}
		\put(0,0)	{\vector(1,4){0.525}}
		\put(0,0)	{\vector(1,2){1}}
		\bezier{20}(1.68,-0.425)(1.68,-0.63)(1.55,-0.77)
		\bezier{20}(0.425,1.68)(0.63,1.68)(0.77,1.55)
		\put(1.2,1.7)	{\makebox(0,0)[l]{$\vec{W}$}}
		\put(-0.2,2)	{\makebox(0,0)[l]{$\vec{W}^{\rm t}$}}
		\put(1.5,-0.55)	{\makebox(0,0){$\theta$}}
		\put(0.55,1.45)	{\makebox(0,0){$\theta$}}
		\end{picture}}
\end{picture}
}
\caption{Geometrical view on generalization.}
\end{figure}
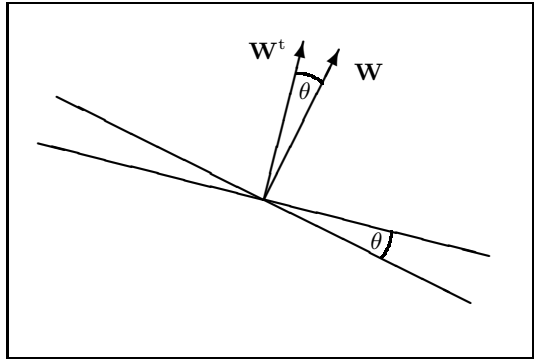

Assume that the student learns the examples according to the generalized Hebb 
rule. In vector notation,
\be
\vW = \frac{1}{N} \sum_{\mu=1}^P \zeta_0^\mu \vxi^\mu\ .
\label{G2}
\ee
An argument in the spirit of the signal-to-noise-ratio analysis used above to
analyse Hebbian learning of random associations can be utilized to obtain 
the generalization error as a function of the size $P$ of the training set. 
To this end, one decomposes each input pattern $\vxi^\mu$ into its contribution
parallel and orthogonal to $\vW^{\rm t}$. Through (\ref{G2}), this decomposition induces a corresponding decomposition of the student's coupling vector, $\vW= \vW_\| + \vW_\perp$. Using (\ref{G1}), one can conclude that the contributions to $\vW_\|$ add up constructively, hence  $\| \vW_\| \|$ grows like $\alpha = P/N$ with the size $P$ of the training set. The orthogonal contribution $\vW_\perp$ to the student's coupling vector, on the other hand, can be 
interpreted as the result of an unbiased $P$--step random walk (a diffusion process) in the $N-1$--dimensional space orthogonal to $\vW^{\rm t}$, each step of length $1/\sqrt N$. So typically $\|\vW_\perp \| \sim  \sqrt \alpha$. In the large system limit, prefactors may be obtained by appeal to the central limit theorem, and the average generalization error is thereby found to be
\be
\varepsilon_{\rm g}(\alpha) = \frac{1}{\pi} \arctan(\|\vW_\perp \| / \| \vW_\| 
\| ) = \frac{1}{\pi} \arctan\la \sqrt{\frac{\pi}{2\alpha}}\ra \ .
\ee
It deacreases from $\varepsilon_{\rm g}\simeq 0.5$ at the beginning of the training session --- the result one would expect for random guesses --- to 
zero, as $\alpha \to \infty$. The asymptotic decrease is $\varepsilon_{\rm 
g}(\alpha) \sim 1/\sqrt{2\pi\alpha}$ for large $\alpha$.

The simple Hebbian learning algorithm is thus able to find the rule asymptotically, although it is never perfect on the training set. A similar 
argument as that given for the generalization error can be invoked to compute
the average training error $\varepsilon_{\rm t}$, which is always bounded from
above by the generalization error. 

How does the perceptron algorithm perform on the problem of learning a rule. 
First, since the examples themselves are generated by a perceptron, hence 
linearly separable, perceptron learning is always perfect on the training set. 
That is $\varepsilon_{\rm t} = 0$ for perceptron learning. To compute the 
generalization error, is not so easy as for the Hebbian student. We shall try
to convey the spirit of such calculations later on in Sec.~\ref{secTL}. Let us
here just quote results.

Asymptotically the generalization error for perceptron learning decreases with
the size of the training set as $\varepsilon_{\rm g}(\alpha) \sim \alpha^{-1}$
for large $\alpha$. The pre\-factor depends on further details. Averaging over 
all perceptrons which do provide a correct classification of the training set,
i.e., over the so--called version space, one obtains $\varepsilon_{\rm g}^{\rm 
av}(\alpha) \sim 0.62/\alpha$. For a student who always is forced to find the 
best separating hyperplane for the training set (its orientation is such that 
the distance of the classified input vectors from either side of the plane is 
{\em maximal\/}) --- this is the so--called optimal perceptron --- one has 
$\varepsilon_{\rm g}^{\rm opt}(\alpha) \sim 0.57/\alpha$. It is known that the 
Bayesian optimal classifier (optimal with repect to generalization rather than 
training) has $\varepsilon_{\rm g}^{\rm Bayes} (\alpha) \sim 0.44/\alpha$, but this classifier itself is {\em not\/} implementable through a simple perceptron.
Extensive discussions of these and related matters can be found in (\cite{gyti};
\cite{wrb}; \cite{opki}; \cite{en}).
 
Thus, perceptron learning generalizes faster than Hebbian learning, however at 
higher `computational cost': the perceptron learner always has to retrain on 
the whole new training set every time a pattern is added to it. A significant 
amount of computational cost is required on top of this, if one always tries 
to find the optimal perceptron.

\subsection{Layered Networks}

To overcome the limitations of simple perceptrons so as to realize input--output
relations more complicated than the linearly separable ones, one may resort to
combining several simple perceptrons to build up more complicated architectures.
An important class comprises the so--called multi--layer networks to which we 
now turn.

In multi--layer networks, the output produced by a single perceptron is not necessarily communicated to the outside world. Rather one imagines a setup where
several perceptrons are arranged in a layered structure, each node in each layer
independently processing information according to its afferent synaptic weights
and its transfer function $\Phi$. The first layer --- the input layer --- 
receives input from external sources, processes it, and relays the processed 
information further through possibly several intermediate so--called hidden  layers. A final layer --- the output layer --- performs a last processing step and transmits the result of the ``neural computation" performed in the layered architecture to the outside world. Synaptic connections are such that no 
feedback loops exist.

Multi--layer networks consisting of {\em simple\/} perceptrons, each 
implementing a linearly separable threshold decision, have been discussed 
already in the early sixties under the name of Gamba perceptrons (see \cite 
{mipa}). For them, no general learning algorithm exists. The situation is 
different, and simpler, in the case where the elementary perceptrons making 
up the layered structure have a smooth, differentiable input--output relation.
For such networks a general--purpose learning algorithm exists, which is 
guaranteed to converge at least locally to a solution, provided that a 
solution exists for the information processing task and the network in 
question.

The algorithm is based on gradient--descent in an ``error--energy 
landscape". Given the information processing task --- a set of input--output
pairs $(\zeta_0^\mu,\vxi^\mu)$, $\mu=1,\dots,P$ to be embedded in the net --- 
and assuming for simplicity a single output unit\footnote{This implies no loss 
of generality. The problem may be analyzed separately for the sub-nets feeding 
each each output node.}, one computes a network error measure over the set of 
patterns
\be
E = \frac{1}{2} \sum_\mu (\delta_0^\mu)^2\ ,
\ee
the output errors $\delta_0^\mu$ being defined as before. For fixed input--output relations $\Phi$, the error measure is determined by the set of 
all weights of the network $E=E(\vW)$. Let $W_{ij}$ be a weight connecting node 
$j$ to $i$. Gradient descent learning aims at reducing $E$ by adapting the 
weights $W_{ij}$ according to
\be
\Delta W_{ij} = - \eta \frac{\partial E(\vW)}{\partial W_{ij}}\ ,
\ee
where $\eta$ is a learning rate that  must be chosen sufficiently small to ensure convergence to (local) minima of $E(\vW)$. For a network consisting of
a single node, one has $\delta_0^\mu = \zeta_0^\mu-\nu_0^\mu$ with $\nu_0^\mu =
\Phi(\sum_j W_{0j} \xi_j^\mu) = \Phi(h_0^\mu)$, hence
\be
\Delta W_{0j} = \eta \sum_\mu \delta_0^\mu \Phi'(h_0^\mu) \xi_j^\mu \equiv \eta 
\sum_\mu \tilde\delta_0^\mu \xi_j^\mu\ ,
\ee
where $\Phi'$ denotes the derivative of $\Phi$. Note that there is a certain similarity with perceptron learning. The change of $W_{0j}$ is related to the product of a (renormalized)  error $\tilde\delta_0^\mu $ at the  output side of
node $0$ with the input information $\xi_j^\mu$, summed over all patterns $\mu$.

If the network architecture is such that no feed--back loops exist, this rule 
is immediately generalized to the multi--layer situation, using the chain rule 
of differential calculus. The resulting algorithm is called the {\em 
back--propagation\/} algorithm for reasons to become clear shortly. Namely,
for an arbitrary coupling $W_{ij}$ in the net one obtains
\be
\Delta W_{ij} = - \eta \sum_\mu \tilde\delta_i^\mu \xi_j^\mu\ ,
\ee
where $\xi_j^\mu$ is the input to node $i$ in pattern $\mu$, coming from node 
$j$ (except when $j$ denotes an external input line, this is not an input from 
the outside world), and $\tilde\delta_i^\mu$ is a renormalized output error at 
node $i$, computed by {\em back--propagating\/} the output--errors of all nodes  $k$ to which node $i$ relays {\em its\/} output via $W_{ki}$,
\be
\tilde\delta_i^\mu = \sum_k \tilde\delta_k^\mu W_{ki} \Phi'(h_i^\mu) \ .
\label{mln1}
\ee
Note that the (renormalized) error is propagated via the link $i\to k$ by utilizing that link in the {\em reverse\/} direction! This kind of error
back--propagation needed for the updating of all links not directly connected
to the output node is clearly biologically implausible. There is currently
no evidence for mechanisms that might provide such functionality in real neural 
tissue. 

Moreover, the algorithm always searches for the nearest local minimum in
the error--energy landscape over the space of couplings, which might be a 
spurious minimum with an untolerably large error measure, and it would be 
stuck there. This kind of malfunctioning of the learnig algorithm can to
some extent be avoided by introducing stochastic elements to the dynamics
which permit occasional uphill--moves. One such mechanism would be provided
by ``online--learning", in which the error--measure is not considered as a
sum of (squared) errors over the full pattern set, but rather as the 
contribution of the pattern currently presented to the net, and by training
on the patterns in some random order.

Back-propagation is a very versatile algorithm, and it is currently the
`work--horse' for training multi--layer networks in practical or technical
applications. The list of real--world problems, where neural networks 
have been successfully put to work, is already rather impressive; see e.g. (\cite{HKP}). Let us just mention two examples. One of the early successes was 
to train neural networks to read (and pronounce) written English text. One
of the harder problems, where neural solutions have recently been found competitive or superior to heuristic engineering solutions, is the prediction
of secondary structure of proteins from their amino-acid sequence. Both
examples share the feature, that algorithmic solutions to these problems are
not known, or at least extremely hard to formulate explicitly. In these, as
in many other practical problems, networks were found to generalize well
in situations which were not part of the training set.

A generally unsolved problem in this context is that of choosing the correct
architecture in terms of numbers of layers and numbers of nodes per layer
necessary to solve a given task. Beyond the fact that a two--layer architecture
is sufficient to implement continuous maps between the input- and output--side,
whereas a three--layer net is necessary, if the map to be realized has 
discontinuities, almost nothing is known (\cite{HKP}). One has to rely on 
trial--and--error schemes along the rule of thumb that networks should be 
{\em as large as necessary, but as small as possible\/}, the first part 
addressing the representability issue, the second the problem that a neural 
architecture that is too rich will not be forced to extract rules from a 
training set but simply memorize the training examples, and so will generalize 
poorly. Algorithmic means to honour this rule of thumb in one way or another ---
under the categories of network--pruning or network--construction 
algorithms --- do, however, exist (\cite{HKP}).

The situation is again somewhat better for certain simplified setups --- 
two--layer Gamba perceptrons where the weights between a hidden layer and
the output node are fixed in advance such that the output node computes a
preassigned boolean function of the outputs of the hidden layer. Popular examples are the so--called committee--machine (the output follows the 
majority of the hidden layer ouputs) and the parity--machine (it produces 
the product of the $\pm 1$ hidden layer outputs). For such machines, storage capacities and generalization curves for random (input) data have been 
computed, and the relevant scales have been identified: The number of random
associations that can be embedded in the net is proportional to the number 
${\cal N}$ of adjustable weights, and in order to achieve generalization, the size $P$ of the training set must also be proportional to ${\cal N}$.
The computations are rather involved and approximations have to be made, which are not in all cases completely under control. Moreover, checks through numerical simulations are hampered by the absence of good learning algorithms. 
So, whereas scales have been identified, prefactors are in some cases still 
under debate. A recent review is (\cite{opki}).  

Neither back--propagation learning (online or off-line) for general multi--layer
networks nor existing proposals for learning in simplified multi--layer 
architectures of the kind just described (see, for instance, the review by 
Watkin et al.~(1993)) can claim a substantial degree of biological plausibility. In this context it is perhaps worth pointing
out a proposal of Bethge et al. (1994), who use the idea of fixing one layer 
of connections the other way round, and consider two--layer architectures
with fixed input--to--hidden layer connections. These provide a preprocessing scheme which recodes the input data, e.g., by representing them locally in terms
of mutually exclusive features. This requires, in general, a large hidden layer
and divergent pathways. The advantage in terms of biological modelling is,
however, twofold. There is some evidence that fixed preprocessing of sensory
data which provides feature detection via divergent neural pathways is found 
in nature, for instance in early vision. Moreover, for learning in the second 
layer, simple perceptron learning can do, which  --- as we have argued above 
--- still has some degree of biological plausibility to it. Quantitative 
analysis reveals that such a setup, one might call it coding--machine, can 
realize mappings outside the linearly separable class (\cite{be+}).
The generalization ability of networks of this type remain to be analyzed
quantitatively. It is clear, though, that the proper scale is again set by
the number of adjustable units.

Interestingly, there exist unsupervised learning mechanisms that can provide
the sort of feature extraction required in the approach of Bethge et al. 
Prominent proposals, which are sufficiently close to biological realism, are 
due to Linsker (1986) and Kohonen (1982; 1989). Linsker suggests a multilayer
architecture of linear units trained via a modified Hebbian learning rule,
for which he demonstrates the spontaneous emergence of synaptic connectivities that create orientation selective cells and so-called center--surround cells
in upper layers, as they are also observed in the early stages of vision. 
Kohonen discusses two--layer architectures where neurons in the second layer 
``compete" for inputs coming from the first, which might be a retina. Lateral 
inhibition, i.e., feedback in the second layer ensures that only a single 
neuron in the second layer is active at a time, namely the one with the largest 
postsynaptic potential for the given input. An unsupervised adaptation process 
of synaptic weights connecting the input layer to the second layer is found to 
generate a system where each neuron in the second layer becomes active for a 
certain group of mutually similar inputs (stimuli). Note that this presupposes 
that similarity of, or correlations between different inputs exists. Inputs
which are mutually similar, but to a smaller degree, excite nearby cells 
in the second layer. That is, one has feature extraction which preserves {\em 
topology\/}. Moreover, the resolution of the feature map becomes spontaneously 
finer for regions of the stimulus space in which stimuli occur more frequently 
than in others. Details can be found in Hertz et al. (1991).

\subsection{A General Theoretical Framework for Analyzing Learning and 
Generalization \label{secTL}}

Let us close the present section with a brief and necessarily very schematic 
outline of a general theoretical framework in terms of which the issues of learning and generalization may be systematically studied. Not because we like
to indulge in formalism, but rather because the theoretical framework itself
adds interesting perspectives to our way of thinking about neural networks in 
general, which, incidentally, carry much further than our mathematical abilities
to actually work through the formalism in all detail for the vast majority of
relevant cases. Key ideas of the approach presented below can be traced back 
to pioneering papers of Elizabeth Gardner (1987; 1988).

To set up the theoretical framework, it is useful to describe the learning 
process in terms of a training energy. Assume that the task put to a network
is to embed a certain set $P$ of input--output pairs $(\vzeta^\mu,\vxi^\mu)$,
$\mu=1,\dots,P$, where the output vectors $\vzeta^\mu$ may be determined from
the input vectors $\vxi^\mu$ according to some rule, or independently chosen.
The training energy may then be written as
\be
E(\vW|\{\vzeta^\mu,\vxi^\mu\}) = \sum_\mu \varepsilon(\vW,\vzeta^\mu,\vxi^\mu)
\ ,
\ee
with a single pattern output error $\varepsilon(\vW,\vzeta^\mu,\vxi^\mu)$ that
is a nonnegative measure of the deviation between the actual network output 
$\vnu^\mu =\vnu(\vW,\vxi^\mu)$ and the desired output $\vzeta^\mu$. In the
case of recursive networks, more specifically, in the case of learning fixed
point attractors in recursive networks, there is of course no need to 
distinguish between input and output patterns.

Learning by gradient descent in an error--energy landscape --- that is learning
as an optimization process --- has been discussed above in connection with the 
back-propagation algorithm for feed--forward architectures, where the absence 
of feedback--loops allowed to obtain rather simple expressions for the 
derivatives of $E$ with respect to the $W_{ij}$. It was noted already in that
context that, in order to avoid getting stuck in local suboptimal energy valleys, one may supplement the gradient dynamics with a source of noise. This would lead to the Langevin dynamics
\be
\Gamma^{-1} \frac{\D}{\D t} W_{ij} = - \frac{\partial}{\partial W_{ij}}
E(\vW | \{\vzeta^\mu,\vxi^\mu\}) + \eta_{ij}(t)\ ,
\label{tl1}
\ee 
in which the (systematic) drift term aims at reducing the training error,
whereas the noise allows occasional moves to the worse. 

There is more to adding noise than its beneficial role in avoiding suboptimal 
solutions. Namely, if the noise in (\ref{tl1}) is taken to be uncorrelated 
Gaussian white noise, with average $\langle \eta_{ij}(t)\rangle = 0$ and 
covariance $\langle \eta_{ij}(t) \eta_{kl} (t') \rangle = 2 T \delta_{(ij),(kl)} \delta(t-t')$, then the Langevin dynamics (\ref{tl1}) is known to converge 
asymptotically to `thermodynamic equilibrium' described by a Gibbs distribution 
over the space of synaptic weights,
\be
\cP(\vW|\{\vzeta^\mu,\vxi^\mu\}) = \cZ^{-1} \exp\{-\beta E(\vW|\{\vzeta^\mu 
,\vxi^\mu\})\}\ .
\label{tl2}
\ee
Here $\beta$ denotes an inverse temperature\footnote{Note that we use temperature $T$ not as specifying ambient temperature, but simply as a measure 
of the degree of stochasticity in the dynamics.} in units of Boltzmann's 
constant, $\beta =1/T$. In the case where the $W_{ij}$ are only allowed to take 
on discrete values, the Langevin dynamics (\ref{tl1}) would have to be replaced
by a Monte--Carlo dynamics at finite temperature, the analog of gradient 
decscent being realized in the limit $T\to 0$. The equilibrium distribution 
would still be given by (\ref{tl2}), if transition probabilities of the 
discrete stochastic dynamics were properly chosen. Note that $\cP$ depends
parametrically on the choice of training examples.

Now two interesting things have happened. First, by introducing a suitable form 
of noise and by considering the long time limit of the ensuing stochastic 
dynamics, we know the distribution $\cP$ over the space of weights explicitly, 
so we can in principle compute averages and fluctuations of {\em all\/} 
observables of which we know how they depend on the $W_{ij}$. Second, by
considering the equilibrium distribution (\ref{tl2}), one is looking at an 
``ensemble of learners" which have reached, e.g., a certain average asymptotic
training error, and one is thereby {\em deemphasizing all details of the 
learning mechanism that may have been put to work to achieve that state}. This 
last circumstance is one of the important sources by which the general 
framework acquires its predictive power, because it is more likely than not 
that we do not know the actual mechanisms at work during learning, and so it 
is gratifying to see that at least asymptotically the theory does not require 
such knowledge.

Of the quantities we are interested in to compute, one is the average training
error
\be
\langle E \rangle =  \langle E(\vW|\{\vzeta^\mu,\vxi^\mu\})\rangle = \int \D 
\mu(\vW)\,\cP(\vW|\{\vzeta^\mu,\vxi^\mu\})\, E(\vW|\{\vzeta^\mu,\vxi^\mu\})\ ,
\ee
where the measure $\D \mu(\vW)$ encodes whatever a--priori constraints might
be known to hold about the $W_{ij}$. It may also be obtained from the ``free
energy" 
\be
\cF = - \beta^{-1} \ln \cZ =  - \beta^{-1} \ln \int \D \mu(\vW) \exp\{-\beta E (\vW|\{\vzeta^\mu,\vxi^\mu\})\}
\label{tl3}
\ee
corresponding to the Gibbs distribution (\ref{tl2}) via the thermodynamic 
relation
\be
\langle E \rangle = \frac{\partial}{\partial \beta} \beta \cF\ .
\label{tl4}
\ee 
The result still depends on the (random) examples chosen for the training set, 
so an extra average over the different possible realizations of the training set must be performed, which gives
\be
E(\beta,P) = \ldd \ld E\rd \rdd = \int \prod_\mu \D \rho(\vzeta^\mu,\vxi^\mu)\,
\ld E\rd\ .
\ee
Such an average is automatically implied, if one replaces the free energy in
the thermodynamic relation (\ref{tl4}) by its average over the possible training
sets, i.e., the so called quenched free energy $\cF_q = - \beta^{-1} \ldd \ln \cZ \rdd$. Similarly, the average generalization error is obtained by first 
considering $\varepsilon_{\rm g}(\vW) = \int \D \rho(\vzeta,\vxi)\,\varepsilon 
(\vW,\vzeta,\vxi)$, that is, the single pattern output error used in (\ref{tl1}), averaged over all possible input output pairs which were not part of the training set, and by computing
\be
\varepsilon_{\rm g}(\beta,P) =\ldd \ld \varepsilon_{\rm g}(\vW) \rd \rdd\ .
\ee
Actually, it turns out that the additional averaging over the various 
realizations of the training set need not really be performed, because each
training set will typically produce the same outcome, which is therefore 
called self--averaging. Technically, however, such averages are usually
easier to handle than specific realizations, and the averages are therefore
nevertheless computed. The same situation is, incidentally, encountered in
the analysis of disordered condensed matter systems. Not too surprisingly 
therefore, it is this subdiscipline of physics from which many of the 
technical tools used in quantitative analyses of neural networks have been 
borrowed.

It is well known that the statistical analysis of conventional condensed matter 
comes up with virtually deterministic relations between macroscopic observables 
characteristic of the systems being investigated, as their size becomes large (think of relations between temperature, pressure and density, i.e., equations
of state for gases). In view of the appearance of relations of statistical 
thermodynamics in the above analysis, one may wonder whether analogous deterministic relations would emerge in the present context. This is indeed
the case, and it may be regarded as the second source of predictive power of
the general approach. 

In the large system limit, that is, as the number $\cN$ of synaptic couplings
becomes large, the distribution (\ref{tl2}) will give virtually all weight to
$\vW$--configurations with the same macroscopic properties. Among these are,
in particular, the training error per pattern, $\varepsilon_{\rm t} = P^{-1} \sum_\mu E(\vW|\{\vzeta^\mu,\vxi^\mu\})$, and the generalization error 
$\varepsilon_{\rm g}$. 

The analysis reveals that a proper large system limit generally requires to 
scale the size $P$ of the training set according to $P = \alpha \cN$, as we 
have observed previously in specific examples. As $\cN\to \infty$ (at fixed 
$\alpha$) learning and generalization errors are {\em typically\/} --- i.e., 
for the overwhelming majority of realizations --- given by their thermodynamic 
averages (as functions on the $\alpha$--scale), $\varepsilon_{\rm t} = P^{-1} 
E(\beta,P)\to \varepsilon_{\rm t}(\beta,\alpha)$ and $\varepsilon_{\rm g}\to \varepsilon_{\rm g}(\beta,\alpha)$. 

The reason for the generalization error to be among the predictable macroscopic 
quantities stems from the fact that it is related to the distance in weight 
space, $\Delta (\vW^{\rm t},\vW) = \cN^{-1} \sum_{ij} (W_{ij}^{\rm t} - 
W_{ij})^2$, between the network configuration $\vW$ and the target configuration $\vW^{\rm t}$ which the learner is trying to approximate. This is itself a 
(normalized) extensive observable which typically acquires non-fluctuating 
values in the thermodynamic limit.

The results obtained via the statistical mechanics approach are, as we have
indicated, {\em typical\/} in the sense that they are likely to be shared by 
the vast majority of realizations. This is to be seen in contrast to a set 
of results about learning and generalization, obtained within the 
machine--learning community under the paradigm of ``probably almost correct 
learning". They usually refer to worst--case scenarios and do, indeed, usually
turn out to be overly pessimistic. We refer to (\cite{wrb}; \cite{en}; 
\cite{opki}) for more details on this matter.

In the zero--temperature ($\beta\to\infty$) limit, the Gibbs distribution 
(\ref{tl2}) gives all weight to the synaptic configurations which realize 
the smallest conceivable training error. An interesting question to study 
in this context is what the largest value of $\alpha$ is, such that the
minimum training energy is still zero. This then gives the size of largest 
pattern set that can be embedded without errors in the given architecture --- 
{\em irrespective\/} of whatever learning algorithm might be used to train
the net. This number is called the absolute capacity of the net, and it 
depends, of course, on the pattern statistics. In the case where outputs in 
the pattern set are generated according to some rule, one obtains information 
as to whether the rule is learnable, i.e., representable in the network under consideration, or not. 

For unbiased binary random patterns, the absolute capacity is found to be 
$\alpha_{\rm c} = 2$ for networks consisting of simple threshold elements, and 
without hidden neurons. The number increases, if the patterns to be embedded 
in the net have unequal proportions of active and inactive bits (see also Sec. 
4 below); it decreases if one wants to embed patterns with a certain stability, that is, such that correct classifications are obtained even with a certain 
amount of distortion at the input side (\cite{gar87}; \cite{gar88}). In 
attractor networks, large stability implies large basins of attraction for the 
patterns embedded in the net. 

Another way to phrase these ideas is to note that learning of patterns puts 
restrictions on the allowed synaptic couplings. The absolute capacity is 
reached when the volume of allowed couplings, which becomes progressively 
smaller, as more and more patterns are being embedded in the net, eventually shrinks to zero. The logarithm of the allowed volume is like an entropy, a
measure of diversity. Learning then reduces the allowed diversity in the
space of (perfect) learners. Similarly, by learning a rule from examples, the 
volume in the space of couplings will shrink with increasing size of the 
training set, and eventually be concentrated around the coupling vector 
representative of the target rule. Generalization ensues. 

An interesting application of these ideas as means to predict the effects
of brain lesions has been put forward by Virasoro (1988). He demonstrated
that after learning hierarchically organized data --- items grouped in classes
of comparatively large similarity within classes, and greater dissimilarity
between classes --- the class information contained in each pattern enjoys a
greater embedding stability than the information that identifies a pattern 
as a specific member of a class. As a consequence, brain lesions that randomly 
destroy or disturb a certain fraction of synapses after learning, will lead to 
the effect that the {\em specific\/} information is lost first, and the class 
information only when destructions become more severe. An example of the 
ensuing kind of malfunctioning is provided by the prosopagnosia syndrome --- 
characterized by the abiltiy of certain persons to recognize faces as faces, 
without being able to distinguish between individual faces. According to all we 
have said before, this kind of malfunctioning must {\em typically\/} be expected
to occur in networks storing hierarchically organized data, when they are being injured. Note moreover that, beyond the fundamental supposition that memory
resides in the synaptic organization of a net, hardly anything else has to be
assumed for this analysis to go through.

It is perhaps worth pointing out that the Gibbs distribution 
(\ref{tl2}) enjoys a distinguished status in the context of maximum--entropy /
minimum--bias ideas (\cite{jay}). It is the maximally unbiased distribution 
of synaptic couplings, subject only to an, at least in principle, observable 
constraint, namely that of giving rise to a certain average training error. 
Together with the notion of concentration of probabilities at entropy maxima
(\cite{jay}), this provides yet another source of predictive power that may
be attributed to the general scheme.

Finally, we should not fail to notice that there is, of course, also room and need for studying learning dynamics proper as opposed to the statistics of 
asymptotic solutions, because information about final statistics tells nothing
about the time needed to reach asymptotia, which is also relevant and 
important information, certainly in technical applications. Here, we leave
it at quoting just one pertinent example. The {\em existence\/} of neural solutions for a given storage task, which may be investigated by considering
the allowed volume in the space of couplings, tells nothing about our ability
to find them. For the perceptron with binary weights, for instance, Horner (1992) has demonstrated that algorithms with a complexity scaling polynomially 
in system size are not likely to find solutions at {\em any\/} non--zero value 
of $\alpha$ in the large system limit, despite the fact that solutions are known to exist up to $\alpha_{\rm c} \simeq 0.83$.


\section{Attractor Networks -- Associative Memory\label{sec4}}

Memory is one of the basic functions of our brain and it also plays a
central role in any computing device. The memory in a computer is
usually organized such that different contents are stored under
different addresses. The address itself, typically a number, has no
relation to the information which is found under its name. The
retrieval of information requires the knowledge of the corresponding
address or additional search engines using key words with lists of
addresses and cross references.

An associative memory is a device which is organized such that part
of the information allows to recall the full information stored. As an
example the scent of a rose or the spoken word `rose' recalls the full
concept {\em rose}, typical forms and colors of its blossoms and leaves,
or events in which a rose has played a role. 

On a more abstract level we would like to have a device in which
certain patterns $\xi^\mu$ are stored and where a certain input $\eta$
recalls the pattern closest to it. This could be achieved by
searching through the whole set of memories, but this would be
rather inefficient.

A neural network is after all a dynamical system. Its dynamics could be
defined by the update rule (\ref{2d}) or equivalently by a set of
nonlinear differential equations
\begin{equation}\label{4a}
\frac{{\rm d} \nu_i(t)}{{\rm d}t}=-\frac1{\bar\tau}\left(\nu_i(t)
-\Phi\Big(\sum_j W_{ij}\nu_j(t)-\vartheta\Big)\right).
\end{equation}
where $\bar\tau$ is some average delay time. It is known from the theory
of dynamical systems that equations of this type have attractors. That
is, any solution with given initial values approaches some small subset
of the full set of available states, which could be a stationary state 
(fixed point), a periodic solution (limit cycle) or a more complicated 
attractor. The set of initial values giving rise to solutions approaching 
the same attractor is called the basin of attraction of this attractor. 
This can now be used to construct an associative memory, if we succeed in 
finding synaptic couplings such that the  patterns to be stored become
attractors. If this is achieved, an initial state not too far
from one of the patterns will evolve towards this pattern (attractor),
provided it was within its basin of attraction.

It is clear that this mechanism requires networks with strong feedback.
In a feed forward layered network with well defined input and
output layers, the information would simply be passed from the input
layer through hidden layers to the output layer, and without input
such a network would be silent. 

The goal is not only to find the appropriate couplings using a
suitable learning rule, but also to estimate how many patterns can be
stored and how wide the basins of attractions are. Wide basins of
attraction are desirable because initial states having a small part in
common with the pattern to be retrieved should be attracted by this
pattern.

\subsection{The Hopfield model}

A great deal of qualitative and quantitative understanding of such
associative memories has come from a model proposed by Hopfield
(\cite{Hop}; \cite{Amit}; \cite{HKP}). Its purpose is to store
uncorrelated binary random patterns
$\xi^\mu_i=\pm1$, where $i=1,\dots ,N$ labels the nodes (neurons) and
\mbox{$\mu=1,\dots ,P$} the patterns to be stored. It employs the modified
Hebb learning rule (\ref{2c})
\begin{equation}\label{4b}
W_{ij}=\frac1N\sum_\mu \xi^\mu_i\xi^\mu_j\ ,
\end{equation}
and one assumes that each node is connected with every other node. For
the dynamics one uses a discretized version of eq. (\ref{4a}), picking a
node $i$ at random and updating its value according to
\begin{equation}\label{4c}
\nu_i(t+\bar\tau)={\rm sgn}\Big(\sum_{j(\ne i)} W_{ij}\nu_j(t)\Big)\ .
\end{equation}

For the analysis of this model it is useful to define an `energy' or
'cost function'
\begin{equation}\label{4d}
E(t)=-\frac{1}{2} \sum_{ij} \nu_i(t)\,W_{ij}\,\nu_j(t)
\end{equation}
for the firing pattern $\nu_i(t)$ at a given time $t$. It can easily be
shown that this function can never increase in the course of time. This
implies that the firing pattern will evolve in such a way that the system 
approaches one of the minima of $E$. This is like moving in a landscape 
with hills and valleys, and going downhill until a local minimum is 
reached. The existence of such a function, called Lyapunov function,  
ensures that the only attractors of such a model are fixed points or in 
the present context stationary firing patterns.

It has to be shown now that, with the above learning rule, the attractors
are indeed the patterns to be stored, or at least close to them. The
arguments are similar to those given in the context of the perceptron.
As measure of the distance between the actual state and a given
pattern we introduce the `overlap'
\begin{equation}\label{4e}
m_\mu(t)=\frac1N\sum_i\xi^\mu_i\,\nu_i(t)
\end{equation}
which is less than or equal to one, and $m_\lambda(t)=1$ signifies that 
the actual firing pattern is that of pattern $\lambda$. If this is the
case, the overlap with all the other patterns will be of order
$1/\sqrt{N}$. Using the overlap, we can write the energy as
\begin{equation}\label{4f}
E(t)=- \frac{1}{2}\sum_\mu m_\mu(t)^2.
\end{equation}
Investigating this in the limit of large $N$, and considering an initial
state such that the initial overlap $m_\lambda(0)$ is the only one
which is of order 1, the remaining ones being of order $1/\sqrt{N}$, 
one may approximate the energy by $E(t)\simeq -m_\lambda(t)^2/2$, assuming 
that $m_\lambda(t)$ remains the only finite overlap for all time. If this
is the case, the energy will decrease and reach its minimal value for
$m_\lambda(t)\to 1$, as $t\to\infty$. That is, the network has reconstructed 
pattern $\lambda$.

For initial states having a finite overlap with more than one pattern,
the attractor reached can be a new state, called spurious state, composed
of parts of several learnt patterns (\cite{AGS}; \cite{Amit}). This tells 
us that the network seems to memorize patterns which have not been learnt. It 
is not clear whether this has to be considered as malfunctioning or whether
it gives room for creativity in the sense of novel combinations of
acquired experience. With a slightly modified dynamics (\cite{Ho87}), a
mixed initial state can also evolve towards the pattern with maximal 
initial overlap. Depending on the overall situation a network might 
switch from one mode to the other.

The picture so far presented holds as long as the loading $\alpha=P/N$ is 
small enough, so that the random contributions to the energy due to the 
$m_\mu(t)\sim1/\sqrt{N}$ with $\mu\ne\lambda$ can be neglected. 

For higher loading, the influence of these remaining patterns  has 
to be taken into account. A more thorough investigation
(\cite{AGS}; \cite{Amit}; \cite{HKP}) shows that this has two effects.
First of all the retrieval states (minima of $E$) are no longer exactly
the learnt patterns, but close to them with a small amount of errors.
For the whole range of loadings for which this kind of memory works, the
final overlap is larger than 0.96, increasing with decreasing loading. In
addition new attractors are created having a small or no overlap with
any of the patterns. Their effect is primarily (\cite{Ho89}) to
narrow the basins of attraction of the learnt patterns. At a critical
loading of $\alpha_c \simeq 0.138$ these states cause a sudden breakdown 
of the whole memory.

This sudden breakdown due to overloading can be avoided by modified
learning rules. Depending on details (see \cite{HKP} section 3) either
the earliest or the most recent memories are kept and the others are
forgotten. It is also possible to keep the earliest and the most recent
memories and to forget those in between, which seems to be the case with
our own memory. Furthermore certain memories can be strengthened or
erased by unconscious events taking place for instance during dream
phases (see \cite{HKP} section 3).

In order to estimate how efficient such a memory works, it is not only
necessary to find out how many patterns can be stored and how many
errors the retrieval states have, it is also necessary to investigate
the size of the basins of attraction, in other words, which amount of a
pattern has to be offered as initial stimulus in order to retrieve this
pattern. An investigation of the retrieval process (\cite{Ho89})
shows that this minimal initial overlap depends on the loading $\alpha$,
and for $\alpha<0.1$ one finds approximately the retrieval condition
$m_\lambda(0)>0.4\,\alpha$. Finally, one can also estimate the gain of
information reached during retrieval. This is the difference between the
information contained in the pattern retrieved and the information that 
must be supplied in the initial stimulus to guarantee successful retrieval. 
This again depends on the loading, and a maximum of 0.1 bit per synapse 
is reached for $\alpha\approx0.12$.

Another quantity of interest is the speed of retrieval. One finds that
almost complete retrieval is reached already after only 3 updates per
node. Inserting numbers for the relevant time scales of neurons one
obtains 30 to 60 msec. This can be compared to measured reaction times
which are typically of the order of 100 to 200 msec.

Apart from other reasons, the Hopfield model is unrealistic in the sense of
requiring complete and symmetric connectivity. The requirement of symmetry 
$W_{ij}=W_{ji}$ ensures, in particular, the existence of an energy or
cost function (\ref{4d}) ruling the dynamics of the network. The
connectivity among cortical neurons is high, of the order of $10^4$
synapses per neuron, but far from being complete, keeping in mind that
already within the range of the dendritic tree of a single neuron
more than $10^5$ other neurons are found. This has been taken into
account in a study (\cite{DGZ}) of a model with randomly diluted
synaptic connections. The overall properties remain unchanged. The
maximal number of patterns is now proportional to the average number $C$ of
afferent synapses per neuron, $P_{\rm max} = \alpha_c C$, with $\alpha_c 
\simeq 0.64$, but the total gain of information per synapse is still similar 
to the value obtained for the original model. A different behavior is found 
as the critical loading, $\alpha_c$, is approached: In this model the basins 
of attraction remain wide, but the number of errors in the retrieval state increases drastically, as $\alpha \to \alpha_c$.

\subsection{Sparse Coding Networks}

As mentioned previously a remarkable feature of cortical neurons is their
low average firing rate. In principle a neuron can produce as many as 300
spikes per second. Recordings on living vertebrate's brains typically
show some cells firing at an elevated rate of up to 30 spikes per second,
but the average rate is much lower, only 1 to 5 spikes per second.
Retaining the proposal of rate coding one has to conclude that typical
firing patterns are sparse in the sense that the number of active neurons
$N_a(t)$ at each time is much less than the number of silent neurons
$N_s(t)$. This means that the mean activity $a(t)=N_a(t)/(N_a(t) + N_s(t))$ 
is low.

Various versions of attractor networks with low activity have been
investigated (\cite{Wil}; \cite{Palm}; \cite{TF}; \cite{AB}) in the
literature. Within the framework of binary McCulloch-Pitts neurons
their state is conveniently represented by $\nu_a=1$ for active
and $\nu_s=0$ for silent neurons. In this case the original Hebb
learning rule (\ref{2b},\ref{2c}) reinforcing the coupling strength
between neurons active at the same time is appropriate.

Obviously this learning rule creates excitatory synaptic connections
only, so in addition inhibitory neurons are required to control the
mean activity of the network, as discussed in section 2. It turns out
that this control has to be faster than the action of the excitatory
synapses. This seems to be supported by the findings that the
connections with inhibitory neurons are short and their synapses are
typically attached to the soma or the innermost parts of the dendrites
of the excitatory pyramidal cells.

The update rule (\ref{4a},\ref{4c}) has to be modified according to the
$(0;1)$ representation using a step function $\Phi(x)=1$ for $x>0$ and
$\Phi(x)=0$ otherwise.

Again such networks can serve as fast associative memories. The maximal
loading depends on the mean activity. It diverges as $\alpha_c\sim
1/a\ln(1/a)$ for \mbox{$a\to0$}. At the same time the information per
pattern decreases with decreasing activity such that the total gain of
information reaches a constant value of 0.72 bit per synapse
(\cite{Ho89}). This value is, however, reached very slowly; for
example, at $a=0.001$ one finds $\alpha_c=30$ and only 0.3 bit information
gain per synapse. Nevertheless, this value exceeds the one found for the 
Hopfield model.

It should be noted that the class of low activity networks just described
only solves the {\em spatial\/} aspect of the low activity issue. However, 
by going one step further and returning to the continuous--time dynamics 
(26), and by using more realistic `graded' neural input--output relations,
one can solve the temporal aspect as well. Neurons which should be firing 
in one of the low activity attractors are then typically found to fire also 
at low rate (\cite{KuB}). Within models of neural networks based on
spiking neurons, this issue has been addressed by Amit et al. (1994; 1996).

One of the virtues of sparse coding networks is in the learning rule. A
change in the synaptic strength is required only if both, the pre- and
the postsynaptic neuron are active at the same time. This implies that
the total number of learning events is reduced compared to a network
with symmetric coding and consequently the requirements on accuracy and
reproducibility of each individual learning process are less stringent.

Another reason why nature has chosen sparse coding could of course be
reduction of energy consumption because each spike requires some extra
energy beyond the energy necessary to keep a neuron alive.

In a sparse coding network it makes sense to talk about the foreground
of a pattern, made up of the active neurons in this pattern, and a
background containing the rest. The foreground is usually denoted as
cell assembly, a notion which goes back to Hebb. The probability that
a neuron belongs to the foreground of a given pattern is
given by the mean activity $a$ of this pattern, which is assumed to be
low. The probability that this neuron belongs simultaneously to the
foreground of two patterns is given by $a^2$. This means that the cell
assemblies belonging to different pattern are almost completely
disjoint. As a consequence mixture states are no problem because their
mean activity is higher and they can be suppressed by the action of the
inhibitory neurons regulating the overall activity. This will play a
role for some of the functions discussed later.

\subsection{Dynamical Attractors}

The attractor network models discussed so far allowed only for fixed point
attractors or stationary patterns as retrieval states. This is a severe
restriction and one can think of many instances where genuine dynamical
attractors are asked for. The reason for the restriction is the
existence of a Lyapunov function which can be traced back to
the symmetry of the couplings $W_{ij}=W_{ji}$. This shows that
asymmetric couplings have to be included if dynamical attractors are to
be constructed (see \cite{HKP}, section 3).

Let us demonstrate this again on a somewhat artificial example. The
desired attractor should be composed of a sequence of patterns
$\xi^\mu_i$ such that pattern $\mu$ is present for some time $\tau$
and then the next pattern $\mu+1$ is presented. The whole set of patterns
with $\mu=1,\dots ,L$ can be closed such that pattern 1 is shown again
after the last pattern $L$ has appeared, generating a periodically
repeating sequence. This is called a limit cycle. The
retrieval of this cycle should work such that the network is
initialized by a firing pattern close to one of the members of the
cycle, say pattern 1, this pattern is completed, and after a time $\tau$
pattern 2 appears and so on.

This can be achieved by using two types of synapses, fast synapses
$W^f_{ij}$ without delay and slow synapses $W^s_{ij}$ with delay
$\tau$. The update rule (\ref{4a}) now reads
\begin{equation}\label{4g}
\frac{{\rm d} \nu_i(t)}{{\rm d}t}=-\frac1{\bar\tau}\left(\nu_i(t)
-\Phi\Big(\sum_j W^f_{ij}\nu_j(t)+\sum_j W^s_{ij}\nu_j(t-\tau)
-\vartheta\Big)\right).
\end{equation}
The appropriate choice of the couplings is (see eq.(\ref{4b}))
\begin{equation}\label{4h}
W^f_{ij}=\frac1N\sum_{\mu=1}^L \xi^\mu_i\,\xi^\mu_j
\qquad\mbox{and}\qquad
W^s_{ij}=\frac{\lambda}{N}\sum_{\mu=1}^L \xi^{\mu+1}_i\,\xi^\mu_j\,
\end{equation}
with pattern $L+1$ being equivalent to pattern 1.

Assume the network was in a random state for $t<0$ and has been brought
into a state close to pattern 1 at $t=0$. For $0<t<\tau$ the slow
asymmetric synapses will have no effect, whereas the fast synapses drive the
state even closer to pattern 1. For $\tau<t<2\tau$ the slow synapses now tend 
to drive the state from pattern 1 to pattern 2, and if they are
stronger than the fast synapses ($\lambda>1$), the state actually
switches to pattern 2, which is then reinforced by the action of the fast
synapses as well. This process is repeated and the whole cycle is
generated. 

Obviously due to the cyclic symmetry any pattern of the cycle
can be used for retrieval. Furthermore it is possible to store more
than one cycle or cycles and fixed points in the same network. For the
storage capacity the total number of patterns in all attractors is
crucial.

The decisive step in this model is the addition of the non--symmetric
slow synapses which ultimately cause the switching between successive
patterns. Devices of this kind have been studied in several variations
(see \cite{HKP}, section 3). 

The mechanism sketched above requires the existence of slow synapses having 
exactly the delay time necessary for the desired timing of the attractor. 
This can easily be relaxed (\cite{Herz}) by assuming a pool of synapses 
$W^\tau_{ij}$ with different delays $\tau$. Employing a modified Hebb 
learning rule (\ref{2b})
\begin{equation}\label{4i}
\Delta W^\tau_{ij}(t)\propto \nu_i(t)\,\nu_j(t-\tau)\ ,
\end{equation}
the training process reinforces specifically those synapses which have
the appropriate delay time and cycles with different times for the
presentation of each individual pattern can be learnt. This learning
rule is actually the natural extension of Hebb's idea, assuming that the
delay is caused primarily by the axonal transmission time.

One can think of other mechanisms to determine the speed at which
consecutive patterns are retrieved. One such mechanism (\cite{HU}) uses
the phenomenon of fatigue or adaptation (see section 2) and some special
properties of sparse coding networks. The process of adaptation can be
mimicked by a time dependent threshold $\vartheta_i(t)$ with
\begin{equation}\label{4j}
\frac{{\rm d}\vartheta_i(t)}{{\rm d}t}
=\frac1{\tau_a}\Big(\vartheta_o+\vartheta'\nu_i(t)-\vartheta_i(t)\Big)
\end{equation}
where $\tau_a$ is the time constant relevant for adaptation. According
to this equation the threshold of a silent neuron relaxes towards
$\vartheta_o$ and is increased if this neuron fires at some finite rate.

For the synaptic couplings again a combination of symmetric
couplings, stabilizing the individual patterns, and non--symmetric
couplings, favoring transitions to the consecutive patterns in the
sequence, is used. This means that eqs.(\ref{4g},\ref{4h}) can again be
used with the above time dependent threshold $\vartheta_i(t)$ but
without retardation in the asymmetric couplings $W^s_{ij}$. In
contrast to the above model, now $\lambda<1$ has to be chosen.

This works as follows. Assume the network was in a completely
silent state for $t<0$ and all the thresholds have their resting value
$\vartheta_o$. Applying an external stimulus exciting the cell
assembly or pool of active neurons of pattern 1, the symmetric couplings
stabilize this pattern. The nodes which should be active in pattern 2
are also excited but if $\lambda$ is sufficiently small the action of the
asymmetric couplings is not strong enough to make them fire, too. As
time goes on, the neurons active in pattern 1 adapt and their threshold
increases, reducing their firing rate. This reduces also the global
inhibition and at some time the action of the weaker asymmetric
couplings will be strong enough to activate the pool of neurons which
have to be firing in pattern 2. This works of course only, if the
neurons of this second pool are still fresh. This is, however, the case
because in a sparse coding network the probability to find a neuron
simultaneously in the cell assemblies of two consecutive patterns is low.
After adaptation of the neurons in the second pool the state switches to
pattern 3 and so on.

\subsection{Segmentation and Binding}

A similar sparse coding network with adaptive neurons can also solve the
problem of segmentation (\cite{HU}; \cite{RGH}). Assume that an external
stimulus excites simultaneously the pools of neurons of more than
one pattern. The task is then to exhibit the separate identity of these 
patterns despite the fact that their representative neuron pools are 
simultaneously excited. This can be achieved by {\em activating}, i.e., 
retrieving only one of the patterns at a time and selecting another one 
a bit later. This is actually what we do, if we are confronted with 
complex situations containing several unrelated objects. We concentrate on 
one object for some time and then go to the next, and so on. 

A sparse coding network with suitable inhibition will allow for the 
activation of a single pattern only, because the simultaneous
recall of two or more patterns would create an enhanced overall
activity which is suppressed by the action of the inhibitory neurons.
If exposed to a stimulus containing more than one learnt pattern, this
network will first activate the pattern having the strongest input.
After some time the pool of active neurons in this pattern will have
adapted and the network retrieves the pattern with the second
strongest stimulus because its pool of neurons is still fresh,
disregarding again the small amount of neurons common to the active
pools of both patterns. This goes on until all patterns contained in the
external stimulus have been retrieved or until the neurons of the first
pool have recovered sufficiently to be excited again. Due to this
recovery only a small number of patterns can be retrieved one after
the other, and those being weakly stimulated will never appear. This is in
accordance with our everyday experience.

This example is of course not a proof that this has to be the way how
segmentation is done in our own brain. It only shows how it could plausibly 
be done. This critique applies, however, to the other models discussed
as well.

A complementary problem is that of binding. Imagine in a visual scene a
large object moves behind some obstacle. What is actually seen is the
front and the back end of this object, with the middle part hidden. The
feature which is common to both parts is the speed at which they move,
and this allows to identify both parts as belonging to one object. If
only one part is moving, they are easily identified as parts of two
different objects. That is, parts of a complex stimulus having
certain features in common are identified as parts of a larger object;
these parts are linked.

A possible mechanism for this linking was discovered in multi--electrode
recordings in the visual cortex of cats or other mammals (\cite{GS};
\cite{Eck}). It was observed that a moving light bar creates an
oscillatory firing pattern in the cells having appropriate receptive
fields. A second light bar created an oscillatory response in some other
neurons. The motion of both bars in the same direction created a
synchronization and phase locking of these oscillations whereas no such
effect was observed if they were moved in different directions. This
effect could even be observed among neurons belonging to different areas
in the visual cortex.

The proposal is now that linking is performed by synchronization of
oscillatory or more general firing patterns. 

The observed oscillations had a period of about 20 msec and lasted for
about 10 periods. Synchrony was established already within the first few
oscillations. It should be pointed out that an individual neuron emits
at most one or two spikes during one period. This means that larger
assemblies of cells with similar receptive fields have to cooperate.
Actually the oscillations were observed in intercellular recordings
which pick up the signals of many adjacent neurons, or in averages over
many runs. The fast synchronization time and the relatively short
duration of the oscillations might indicate that the important feature
is not so much the existence of these oscillations, but rather the
synchronous activity within a range of a few msec.

Not too surprisingly several idealized models have been proposed
reproducing this effect. Most of them are still based on a rate coding
picture. This seems problematic in view of the short times involved and
the relatively low average spiking rates of any individual neuron.
Nevertheless rate coding is not completely ruled out, if one keeps in 
mind that a rate has to be understood not as a temporal average over a single
cell but rather as an average over assemblies of similar cells.

\subsection{Synchronization of Spikes and Synfire Chains}

Rate coding is the widely accepted paradigm for the predominant part
of data processing in the brain of vertebrates. Keeping in mind that
rates might have to be understood as averages over groups of neurons,
elementary operations could be performed within the integration time of
a neuron, typically 10 msec. The exact timing of the incoming
spikes within this period should not matter.

If, on the other hand, a short volley of synchronized spikes arrives at a
neuron within a fraction of a msec, this neuron can fire within a
fraction of a msec. This can be used for very fast data processing
whenever necessary, for instance in the auditory pathway where phase
differences in the signals coming from the two ears are analyzed.

This raises the issue whether such short volleys of synchronized spikes
are a general feature, and what new kind of data processing can be made
this way.

A possible such mechanism are synfire chains (\cite{Abe}). Their
building blocks are pools of neurons locally connected in a feed forward 
manner. If the neurons in one pool are stimulated simultaneously, they 
will emit synchronized spikes. After some short delay time these spikes
arrive at the neurons forming the next pool and cause a synchronous
firing of this pool too. This process is repeated and a wave of
activity travels with a certain speed along the chain. Actually the
neurons forming the pools are all members of a larger network and a
given neuron can belong to several pools. The chain and its pools are
only defined by their connectivity. There might be also connections from
the neurons of one pool to other neurons not belonging to the next pool.
These connections have to be weak, however, otherwise those postsynaptic
neurons have to be counted as members of the next pool. The picture of
distinct pools is somewhat washed out if variations in the delay times
are taken into account. What matters is the synchronous timing of the
incoming spikes.

In some sense the idea of synfire chains is closely related to the
dynamic attractors discussed earlier. The difference is in the sharp
synchronization of the volleys of spikes. Model calculations show that
some initial jitter in the volleys can even be reduced and synchrony 
sharpened up, stabilizing the propagation along the chain.

What would be the signature of synfire chains as seen in multi--electrode
recordings of the spike activity? In such recordings the spikes emitted
by few neurons picked at random are registered. If a synfire chain is
triggered a certain temporal spiking pattern should be generated
depending on where in the chain those neurons are located. If this
synfire chain is active repeatedly, the spiking pattern should also
repeat and the corresponding correlations should become visible
against some background activity. Apparently such correlations have
been observed with spiking patterns extending over several hundred msec
and with a reproducibility of less than one msec (\cite{Abe94}). This is
quite remarkable, and it requires that a sufficient number of neurons is
involved such that irregularities in the precise timing of the
individual spikes are averaged out.

If the total number of neurons involved in a synfire chain or in one of
its pools is small compared to the size of the total network, it is
possible that several synfire chains are active at the same time.
Assuming a weak coupling between different chains, synchronization of
chains representing different features of the same object could be of
relevance for the binding problem.

On the other hand simulations on randomly connected networks with
spiking neurons and low mean activity show the existence of transients
and attractors resembling synfire chains. What is typically found is a
small number of long limit cycles and in addition a small number of
branched long dominating transients leading into the cycles. An
arbitrary initial state is quickly attracted to one of the pronounced
transients or directly to one of the limit cycles. The emerging picture
resembles a landscape with river systems (transients) and lake--shores
(cycles). It is possible that these structures serve as seeds for more
pronounced synfire chains formed later by learning. It is also possible
that synfire type activity is just a byproduct of other data processing
events or of background activity, if such transients and attractors are
always present and are not erased by learning. 


\section{Epilogue}

The present contribution has been concerned with investigations of neural
networks {\em as information processing devices}. The basic assumption that
has been underlying these investigations is that information is represented 
by neural firing patterns, and that the spatio--temporal evolution of these
patterns is a manifestation of information processing. Its course is determined
by the synaptic organization of a net, which can itself evolve on larger time
scales through learning. Neural networks are thus dynamical systems on (at 
least) two levels --- that of the neurons and that of the synapses.

For higher vertebrates, there is some evidence that both, speed and reliability
of neural `computations' are achieved by being performed in {\em large\/}
networks employing a high degree of parallelism. It makes up for the relatively 
slow dynamics of single neurons, and it gives rise to a remarkable robustness
of network--based computation against malfunctioning of individual neurons or
synaptic connections.

The fact that we are dealing with large systems, when we are trying to 
understand neural information processing, indicates that concepts of statistical physics might provide useful tools to use in such an endeavour. This proves, 
indeed, to be the case, again on (at least) two levels --- for the analysis of 
{\em neural\/} dynamics and associative memory,  and for the analysis of the 
{\em synaptic\/} dynamics associated with learning and generalization.

The robustness of neural information processing against various, even rather severe kinds of malfunctioning at a microscopic level --- mentioned above as an 
observational fact --- shows that microscopic details may be varied in such systems without necessarily changing their overall properties. This is to be 
seen as a {\em hint\/} that even rather simplified models might capture the {\em essence\/} of certain information processing mechanisms whithout necessarily 
being faithful in the description of all details. 

Conversely, the analysis of simplified models {\em reveals\/} that information
processing in neural networks {\em is\/} robust against changing details at
the microscopic level, be they systematic or random. For example, the main
feature of the Hopfield model (1982), viz. to provide a mechanism for 
associative information retrieval at moderate levels of loading, has been 
found to be insensitive against a wide spectrum of variations affecting 
virtually all characteristics of the original setup --- variations concerning neural dynamics, learning rules, representation of neural states, pattern 
statistics, synaptic symmetry, and more. Similarly, the ability of neural 
networks to acquire information through learning and to generalize from examples was observed to be resilient against a large variety of modifications of the learning mechanism.

We should not fail to point out once more that the statistical approach to 
neural networks can claim strength and predictive power {\em only\/} in the 
description of macroscopic phenomena emerging as cooperative effects due to 
the interaction of many neurons, either in unstructured or in homogeneously 
structured networks. We have indicated that, indeed, a number of interesting 
information processing capabilities belong to this category. Our ability to 
analyze them quantitatively has been intimately related to finding the proper 
macroscopic level of description, which by itself is almost tantamount to 
finding the proper questions to be addressed in understanding various brain functions.

In concentrating on specific brain functions, mechanisms and processes 
realizable in specific unstructured or homogeneously structured architectures, 
we had to leave untouched the question of how these various functions and 
processes are being put to work simultaneously in a real brain --- supporting each other, complementing each other, and communicating with each other in the most intricate fashion. A central nervous system is after all {\em not\/} an 
unstructured or homogeneously structured object, but rather exhibits rich 
structures on many levels, with and without feedback, with and without 
hierarchical elements. Analyzing the full orchestration of neural processes in 
this richly structured system is currently way beyond our capabilities --- not 
in small part perhaps due to the fact that we have not yet been able to discover the proper way of looking at the system as a whole.

Whether, in particular, the emergence of the `self' will eventually be 
understood through and as an orchestration of {\em neural processes}, we cannot 
know. In view of the richness of phenomena we have observed already at the level of simple, even primitive systems we see, however, no strong reason to exclude this possibility.


\end{document}